\documentclass{article}

\usepackage[preprint]{neurips_2021}

\usepackage[utf8]{inputenc} 
\usepackage[T1]{fontenc}    
\usepackage{hyperref}       
\usepackage{url}            
\usepackage{booktabs}       
\usepackage{amsfonts}       
\usepackage{amsmath}
\usepackage{mathtools}
\usepackage{nicefrac}       
\usepackage{microtype}      
\usepackage{xcolor}         
\usepackage{multirow}
\usepackage{graphicx}
\usepackage{bm}
\usepackage{rotating}

\title{PlasmoFAB: A Benchmark to Foster Machine Learning for \textit{Plasmodium falciparum} Protein Antigen Candidate Prediction}

\author{%
    \hspace{1mm}Jonas C.~Ditz \\
    \textit{Shared First Author}\\
	Methods in Medical Informatics\\
	Department of Computer Science\\
	University of T\"{u}bingen\\
	T\"{u}bingen, Germany \\
	\texttt{jonas.ditz@uni-tuebingen.de} \\
	\And
	\hspace{1mm}Jacqueline Wistuba-Hamprecht \\
    \textit{Shared First Author}\\
	Methods in Medical Informatics\\
	Department of Computer Science\\
	University of T\"{u}bingen\\
	T\"{u}bingen, Germany \\
	\small{\texttt{jacqueline.wistuba-hamprecht@uni-tuebingen.de}} \\
    \AND
	\hspace{1mm}Timo Maier \\
	Computomics GmbH\\
	T\"{u}bingen, Germany\\
    \And
	\hspace{1mm}Rolf Fendel \\
	Institute of Tropical Medicine\\
	University Hospital T\"{u}bingen\\
	T\"{u}bingen, Germany \\
	\texttt{rolf.fendel@uni-tuebingen.de} \\
	\And
	\hspace{1mm}Nico Pfeifer \\
	Methods in Medical Informatics\\
	Department of Computer Science\\
	University of T\"{u}bingen\\
	T\"{u}bingen, Germany \\
	\texttt{nico.pfeifer@uni-tuebingen.de} \\
    \And
	\hspace{1mm}Bernhard Reuter \\
	Methods in Medical Informatics\\
	Department of Computer Science\\
	University of T\"{u}bingen\\
	T\"{u}bingen, Germany \\
	\texttt{bernhard.reuter@uni-tuebingen.de} \\
}

\begin{document}

\maketitle

\begin{abstract}
  \textbf{Motivation:} Machine learning methods can be used to support scientific discovery in healthcare-related research fields. However, these methods can only be reliably used if they can be trained on high-quality and curated datasets. Currently, no such dataset for the exploration of \textit{Plasmodium falciparum} protein antigen candidates exists. The parasite \textit{Plasmodium falciparum} causes the infectious disease malaria. Thus, identifying potential antigens is of utmost importance for the development of antimalarial drugs and vaccines. Since exploring antigen candidates experimentally is an expensive and time-consuming process, applying machine learning methods to support this process has the potential to accelerate the development of drugs and vaccines, which are needed for fighting and controlling malaria.\\
  \textbf{Results:} We developed \textit{PlasmoFAB}, a curated benchmark that can be used to train machine learning methods for the exploration of \textit{Plasmodium falciparum} protein antigen candidates. We combined an extensive literature search with domain expertise to create high-quality labels for \textit{Plasmodium falciparum} specific proteins that distinguish between antigen candidates and intracellular proteins. Additionally, we used our benchmark to compare different well-known prediction models and available protein localization prediction services on the task of identifying protein antigen candidates. We show that available general-purpose services are unable to provide sufficient performance on identifying protein antigen candidates and are outperformed by our models that were trained on this tailored data.\\
  \textbf{Availability:} \textit{PlasmoFAB} is publicly available on Zenodo with DOI 10.5281/zenodo.7433087. Furthermore, all scripts that were used in the creation of \textit{PlasmoFAB} and the training and evaluation of machine learning models are open source and publicly available on GitHub here: \href{https://github.com/msmdev/PlasmoFAB}{https://github.com/msmdev/PlasmoFAB}.
\end{abstract}

\section{Introduction}
Malaria is a major health problem worldwide, causing more than 247 million cases and approximately 619,000 deaths in 2021 {\citep{world2022world}}. Almost all malaria cases are caused by \textit{Plasmodium falciparum} (Pf), predominantly in Africa. Children, pregnant women, and malaria-na\"\i ve subjects are at high risk to develop severe malaria {\citep{riley2013immune,wu2019evaluation}}. Furthermore, the increase in resistance to both insecticides that target the mosquito vector and anti-malaria drugs, as well as the COVID-19 pandemic, led to an increase of morbidity in several highly endemic countries in the past years {\citep{WHO2021}}. Vaccines are very effective means in protecting against infectious diseases as recently demonstrated in the case of COVID-19. The RTS,S vaccine is the first malaria vaccine recommended by the World Health Organization (WHO) for widespread use in children in endemic settings with a substantial reduction of severe malaria cases, but limited reduction of transmission of malaria{\citep{rts2015efficacy,olotu2013four}}. Besides this first success in fighting severe malaria, there is still an urgent need to develop an effective malaria vaccine that confers sterile protection and reduces malaria transmission. However, developing an effective malaria vaccine is still challenging due to the complex, multi-stage life-cycle of Pf, which is genetically highly diverse and employs several immune evasion strategies. As a result, our understanding of immune responses to Pf-specific antigens that mediate naturally acquired or experimentally induced protection is incomplete. 

More than 5,300 genes are expressed during the life-cycle of Pf {\citep{obiero2019antibody}}. However, only a small subset of proteins that are expressed by Pf is considered in current target candidate screening processes for an effective malaria vaccine {\citep{jagannathan2022malaria,mordmuller2017sterile}}. Since most of the unused proteins have unknown function and experimental validation remains costly and time-intensive, computational methods can be used for pre-screening of proteins of interest. For example, trans-membrane topology prediction is an established task in bioinformatics, where the aim is to predict how and if a protein resides in the cell membrane, i.e., predict the location and length of trans-membrane domains. The class of membrane proteins is one of the most important classes of proteins for medical use. About 25-30\% of natural proteins reside in the cell membrane and are, thus, often bound by antibodies during an immune reaction {\citep{baker2010making}}. Another class of relevant proteins for vaccine and drug development is the class of exported proteins. Many of these fulfill important functions for parasite survival. For example, certain proteins ensure that infected red blood cells stick to the microvasculature, one of the factors that makes malaria a potentially fatal disease {\citep{tuteja2007malaria,wahlgren2017variant}}. In recent years, several scholars developed general-purpose models for sub-cellular localization prediction and offered them as prediction services to be used by the academic community {\citep{krogh2001tmhmm,hallgren2022deeptmhmm,almagro2017deeploc,thumuluri2022deeploc,kall2007advantages}}. While general-purpose models provide researchers with an easy-to-use solution for performing prediction tasks, the lack of out-of-distribution generalization capabilities of most general-purpose models leads to sub-optimal prediction performances on novel datasets and misleading pre-screening results {\citep{ye2021towards}}. However, training supervised machine learning models for protein antigen candidate prediction needs a sufficient amount of protein sequences with high-quality labels. Currently, only a small fraction of publicly available Pf protein sequences have high-quality labels, making the training of models for identification of such antigens for vaccine and drug development exponentially harder. With this work, we introduce the \textit{\textbf{Plasmo}dium \textbf{F}alciparum-specific \textbf{A}ntigen candidate \textbf{B}enchmark} (\textit{PlasmoFAB}), a manually pre-processed and curated dataset containing labeled protein sequences for \textit{Plasmodium falciparum} protein antigen candidate prediction.

This manuscript is structured as follows. We describe in detail the process of creating \textit{PlasmoFAB} including the used data sources, pre-processing, and validation steps. Afterwards we present our experiments for predicting Pf protein antigen candidates. Here we show the limitations of using established tools and present approaches that provide solutions to overcome these limitations. We conclude our work with a discussion about necessary actions that have to be taken in order to further improve \textit{PlasmoFAB} and, hence, further foster the development of vaccines and drugs to control malaria.

\section{\textit{PlasmoFAB}: Plasmodium Falciparum-specific Protein Antigen Candidate Benchmark}
\label{sec:plasmofab}

\begin{figure}[t]
    \centering
    \includegraphics[width=\textwidth]{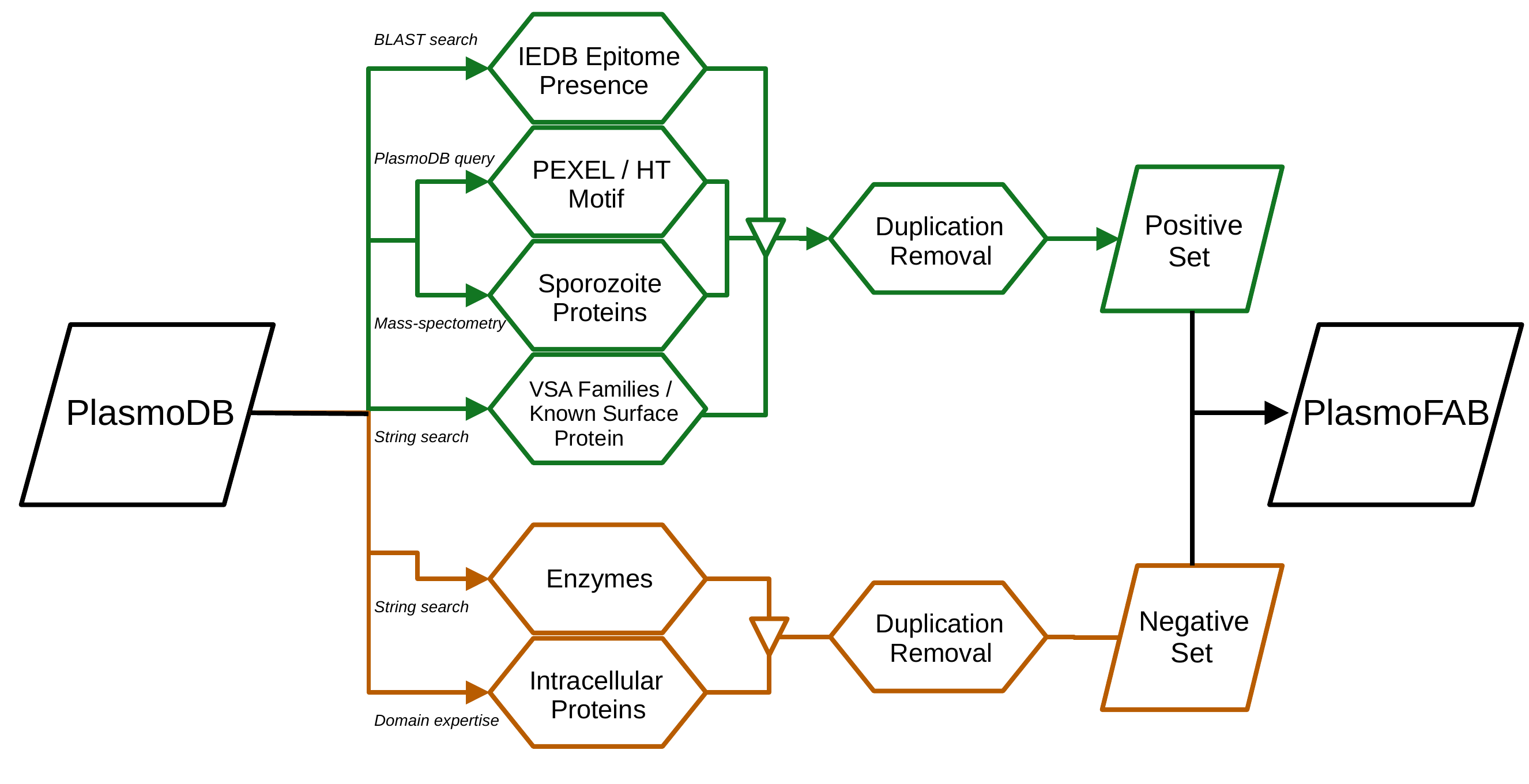}
    \caption{Schematic overview of the pre-processing steps for the creation of \textit{PlasmoFAB}. The green workflow shows the pre-processing of the positive set, i.e., Pf proteins that are either extracellular or membrane-located which renders them eligible to be considered as antigen candidates. We used knowledge-driven techniques like algorithmic homology search, mass-spectrometry, string search, and validation by published literature to create sets of proteins containing antigen candidates. These sets were merged and duplicates were removed to create the positive set. The red workflow shows the pre-processing of the negative set. We combined enzymes validated by published literature with proteins that were assigned to be intracellular by a domain expert and UniProtKB/SwissProt (reviewed) to create the negative set.}
    \label{fig:preprocessing}
\end{figure}

\begin{table}[]
    \centering
    \caption{Composition of the \textit{PlasmoFAB} benchmark. The difference between the sum of sequences in each inclusion criterion and the total number of unique sequences in the positive set occurs due to the fact that some proteins fulfill more than one inclusion criterion. These proteins were not duplicated, resulting in the mismatch between the sum of proteins in each criterion and the total number of proteins in \textit{PlasmoFAB}.}
    \tiny
    \hspace*{-1.2cm}
    \begin{tabular}{lllrlll}
        \toprule
        \multicolumn{3}{c}{\textbf{Positive Set (Unique Total = 438)}} & \phantom{abc} & \multicolumn{3}{c}{\textbf{Negative Set (Unique Total = 384)}} \\
        \cmidrule{1-3} \cmidrule{5-7} 
        Inclusion Criterion & Identified By & \# Proteins & \phantom{abc} & Inclusion Criterion & Identified By & \# Proteins \\
        \midrule
        IEDB epitope & BLAST match (high confidence) & 57 & \phantom{abc} & Intracellular Proteins & \multirow{2}{*}{\shortstack[l]{Combined string and literature search; \\ Domain expertise}} & 384 \\
        IEDB epitope & BLAST match (medium confidence) & 60 & \phantom{abc} &  &  &  \\
        PEXEL/HT motif & \textit{PlasmoDB} query & 265 & \phantom{abc} &   &  &  \\
        Sporozoite proteins & Mass-spectrometry {\citep{swearingen2016interrogating}} & 13 & \phantom{abc} &   &  &  \\
        VSA family / Membrane proteins & Combined string and literature search & 302 & \phantom{abc} &   &  &  \\
        \bottomrule
    \end{tabular}
    \label{tab:pfal_stats}
\end{table}

The term supervised machine learning (SL or supervised ML) summarizes techniques that correlate patterns within datasets to desired output variables, i.e., labels for classification or continuous values for regression. The foundation of using supervised ML methods for scientific discovery in medical research are curated datasets with validated and biologically meaningful output variables. Currently, there is no benchmark that fulfills this prerequisite for the exploration of Pf protein antigen candidates. With this manuscript, we tackle this fundamental obstacle for supporting Pf protein antigen candidate exploration with supervised ML techniques.

In the humoral immune response, the production of antibodies is an important step in getting rid of  pathogens. To enable this response chain, pathogen-specific antigens activate B-cells and their differentiation into antibody secreting plasma cells. Therefore, an antigen candidate has to be visible by the humoral immune system of the host. Pf protein antigen candidates can be considered visible, if they are present on the outside of infected host cells, like surface proteins, transmembrane proteins, membrane-located proteins or exported proteins. The VEuPathDB database \textit{PlasmoDB} {\citep{amos2022veupathdb}} contains the complete genome of different Plasmodium species. The protein sequences of the reference strain 3D7 of Pf, available in \textit{PlasmoDB}, are the data source for our curated benchmark. We only selected sequences with experimental evidence, i.e., the corresponding Pf protein has to be referenced in published work with a unique publication identifier. However, these sequences do not have a sub-cellular location label. We combined an extensive literature search with domain expertise to create high-quality sub-cellular location labels that can be used to train ML models on the task of protein antigen candidate prediction for Pf. In other words, \textit{PlasmoFAB}'s positive set contains Pf proteins that are accessible at the surface or the exterior of infected cells, like surface proteins, transmembrane proteins, membrane-located proteins, or exported proteins. On the other hand, \textit{PlasmoFAB}'s negative set contains intracellular proteins, which are needed by the parasite to maintain the intracellular life cycle in hepatocytes or erythrocytes. The executed pre-processing steps for the creation of \textit{PlasmoFAB} are detailed in the following section. A schematic overview of our pre-processing can be found in Figure \ref{fig:preprocessing} and the basic statistics of \textit{PlasmoFAB} are shown in Table \ref{tab:pfal_stats}.

\subsection{IEDB Epitopes}
\label{sec:epitopes}
An epitope is the part of an antigen that is recognized by the immune system of a host organism, i.e., the binding site of an antibody. The Immune Epitope Database (IEDB, \href{https://www.iedb.org/}{https://www.iedb.org/}, \cite{vita2019immune}) contains sequences of known epitopes. We used exact string matching and BLAST similarity matching to compare Pf protein sequences with sequences contained in the IEDB. Proteins that either contained exact matches of epitope sequences or a positive BLAST hit with high or medium confidence score were labeled as antigen candidates for our benchmark.

\subsection{PEXEL/HT Motif}
The majority of Pf proteins that are either exported into the extracellular space by the parasite or integrated into the membrane of infected erythrocytes contain a specific amino acid sequence called Plasmodium exported element (PEXEL) or host targeting (HT)  {\citep{osborne2010host,jonsdottir2021defining}}. Therefore, the presence of this motif is a strong indicator of a protein antigen candidate. \textit{PlasmoDB} indicates the presence of the PEXEL/HT motif within a sequence by a flag in one of its data fields. For \textit{PlasmoFAB}, we included all proteins with the PEXEL/HT motif as positive antigen candidates.

\subsection{VSA families and known membrane proteins}
Variant surface antigen (VSA) families describe proteins that are typically located on cell surfaces. There are three known VSA families in the Pf genome: Plasmodium falciparum erythrocyte membrane protein 1 (PfEMP1), repetitive interspersed family (RIFIN), and sub-telomeric variable open reading frame (STEVOR). The first family, PfEMP1, summarizes proteins that are expressed on the surface of infected erythrocytes during the trophozoite and schizont stage of the infection cycle. These proteins are mainly responsible for effective evasion of immune responses {\citep{wahlgren2017variant}}. Proteins belonging to the RIFIN family are exported onto the cell surface of infected erythrocytes as well. They mediate the sequestration of erythrocytes which results in erythrocyte rosetting that further helps parasites to evade immune responses and can block the blood flow {\citep{wahlgren2017variant}}. Similar to the other two VSA families, STEVOR proteins are also used by Pf parasites to evade host immune responses. They play active roles in the trophozoite, schizont, merozoite, and gametocyte stages of the infection cycle {\citep{gardner2002genome,wahlgren2017variant}}. Beside the members of VSA families, there are a number of known membrane proteins. In the sporozoite stage, those include thrombospondin-related anonymous protein (TRAP), also known as sporozoite surface protein 2 (SSP2), apical membrane antigen 1 (AMA1), liver stage antigen 1 (LSA1), and exported protein 1 (Exp-1), also known as circumsporozoite-related antigen (CRA). Additionally, we included known surface proteins that can be found in other stages of the infection cycle like the family of monomeric serine-threonine protein kinases (FIKK, \cite{ak2021plasmodium}), the helical intersperse sub-telomeric family of exported proteins (PHIST, \cite{tarr2014conserved}), and the multigene family of cytoadherence linked asexual gene (CLAG, \cite{gupta2015conserved}).

Each entry in \textit{PlasmoDB} has a textual product description field containing information about the sample in textual form. We performed a string search on the textual product description field using the names of the VSA families as search terms: '\textit{PfEMP1}', '\textit{RIFIN}', '\textit{STEVOR}'. For additional known membrane and exported proteins, we did not only included the names but also descriptive search terms since the textual product description field is not standardized. The additional search terms were '\textit{surface}', '\textit{circumsporozoite}', '\textit{membrane}', '\textit{exported}', '\textit{serine repeat antigen}', '\textit{TRAP}', '\textit{FIKK}', '\textit{GLURP}', '\textit{CLAG}', '\textit{PHIST}', and '\textit{GPI-anchor}'. However, the source and rationale behind the annotation in \textit{PlasmoDB}'s textual product description field are not always disclosed. To ensure that only validated membrane and exported proteins are included in our benchmark, we performed a literature search for each protein that was selected by our string search and included only proteins with published experimental evidence into our benchmark. To further enrich the set of known membrane proteins, we added a list of sequences validated by the UniProtKB/SwissProt (reviewed) database. This database contains high quality, manually annotated proteins sequences {\citep{uniprot2023uniprot}}.

\subsection{Sporozoite surface-exposed proteins}
The authors in {\citep{swearingen2016interrogating}} used mass-spectrometry to identify potential surface-exposed sporozoite proteins of Pf. They assigned priority scores to each investigated protein ranging from 1 (high confidence) to 6 (low confidence). We downloaded the publicly available data from {\citep{swearingen2016interrogating}} and selected all proteins with a priority score from 1 to 3. We used the unique transcript ID of these proteins to merge this information into the \textit{PlasmoDB} data table and included them into our benchmark as antigen candidates.

\subsection{Intracellular proteins}
The pre-processing steps described above added positive samples, i.e., Pf protein antigen candidates, to our benchmark. However, \textit{PlasmoFAB} needs negative samples, i.e., proteins that are not Pf protein antigen candidates, to be usable for training of supervised ML methods. A model can only learn to detect true protein antigen candiates, if a set of high-quality negative samples, a so-called negative set, is available. Similar to the positive samples, we curated the negative samples to ensure that only intracellular Pf proteins are included into the negative set. Intracellular proteins can only leave the cytoplasm in specific situations that do not reliably occur in the infection cycle, like the burst of an infected erythrocyte or if macrophages digest an infected erythrocyte and subsequently present an intracellular protein as an antigen. However due to the unreliability of these incidents and the fact that both can only occur late in the infection cycle, intracellular proteins are not suitable as antibody targets. Enzymes constitute a subset of intracellular proteins. We performed a string search with the term '\textit{*ase}' on \textit{PlasmoDB}'s textual product description field and included all proteins with published experimental evidence of being enzymes into the negative set of our benchmark. While there is a small number of enzymes that are exported to the cell membrane, we made sure to exclude all enzymes from the negative set for which published experimental evidence of being membrane-located exists. Furthermore, we included a list of known intracellular proteins compiled by a domain expert and a list of intracellular proteins validated by UniProtKB/SwissProt (reviewed).

\section{Utilizing Machine Learning for Plasmodium Falciparum Protein Antigen Candidate Exploration}
Manually exploring Pf proteins for potential antigen candidates is a time consuming and expensive procedure. With the help of our curated benchmark, we can utilize supervised ML to accelerate the process with a pre-screening of potential proteins that reduces the required workload of researchers in the laboratory. The usefulness of such a pre-screening process highly depends on the accuracy that prediction models are able to achieve. We compared the performance of several ML approaches that are commonly used for textual data, especially for biological sequences. The used methods include a kernelized support vector machine (SVM) utilizing the oligo kernel {\citep{meinicke2004oligo}}, the protein language model embedding ESM-1b {\citep{rives2021biological}} combined with a logistic regression (LR) classifier as well as an SVM, and the protein language model embedding ProtT5 {\citep{elnaggar2020prottrans}}, which we also combined with an LR classifier and an SVM. Furthermore, we also tested the performance of existing protein localization prediction tools on the Pf protein antigen candidate prediction task. These tools are publicly offered as a service for protein localization prediction tasks and included TMHMM {\citep{krogh2001predicting}}, DeepTMHMM {\citep{hallgren2022deeptmhmm}}, DeepLoc 1.0 {\citep{almagro2017deeploc}}, DeepLoc 2.0 {\citep{thumuluri2022deeploc}}, and Phobius {\citep{kall2007advantages}}. To ensure a fair comparison between pre-trained prediction services and our self-trained models, we defined a test set that was separated from the training data before model training was performed. We used \textit{MMseqs2} {\citep{zimmermann2018completely,gabler2020protein}} to ensure that each sequence in the test set had at most 30\% homology to sequences in the training set, which is the default setting of \textit{MMseqs2}. The test set consists of 60 sequences (30 antigen targets and 30 intracellular proteins) with the remaining 788 sequences in \textit{PlasmoFAB} used as a training set. All performance measures shown in this section are computed on the test set.

To assess the performance of each method, we used three performance measures that are widely used in computational biology due to their ability to handle imbalanced data with relative ease. First, we used balanced accuracy, which has different definitions in literature. We used the arithmetic mean of sensitivity and specificity {\citep{pedregosa2011scikit}} given by
\begin{equation}
    \label{eq:balancedAcc}
    \text{Acc}_{\text{bal}} = \frac{1}{2} \left( \frac{\text{TP}}{\text{TP} + \text{FN}} + \frac{\text{TN}}{\text{TN} + \text{FP}} \right),
\end{equation}
where TP is the number of correctly predicted protein antigen candidates (i.e., true positives), FP is the number of wrongly predicted protein antigen candidates (i.e., false positives), TN is the number of correctly predicted intracellular proteins (i.e., true negatives), and FN is the number of wrongly predicted intracellular proteins (i.e., false negatives). Additionally, we used the F$_1$-score that is the harmonic mean of precision and recall {\citep{taha2015metrics}} given by
\begin{equation}
    \label{eq:f1}
    \text{F}_1 = \frac{2 \text{TP}}{2 \text{TP} + \text{FP} + \text{FN}},
\end{equation}
with TP, FP, and FN defined in the same way as above. Finally, we also included the Matthews correlation coefficient (MCC, \cite{chicco2020advantages}), which is widely recognized as one of the most reliable performance measures for binary classification on biological data. The MCC is defined as
\begin{equation}
    \label{eq:mcc}
    \text{MCC} = \frac{\text{TP} \cdot \text{TN} - \text{FP} \cdot \text{FN}}{\sqrt{(\text{TP} + \text{FP}) (\text{TP} + \text{FN}) (\text{TN} + \text{FP}) (\text{TN} + \text{FN})}}.
\end{equation}
Again, the definition of TP, FP, TN, and FN are the same as above. Since the classes in \textit{PlasmoFAB} are balanced, we also report precision, recall, and specificity to provide a quick overview over the distribution of FN and FP for the predictions of the tested models.

\subsection{Using \textit{PlasmoFAB}'s training sequences for model training}
Hyperparameter optimization and model selection was exclusively performed on \textit{PlasmoFAB}'s training sequences to avoid information leakage from the test sequences. As a baseline model, we trained a kernelized SVM utilizing the oligo kernel, a kernel function that was specifically developed for biological sequences {\citep{meinicke2004oligo}}. This kernel computes the similarity of two sequences based on $k$-mer occurrence with a tunable degree of positional uncertainty. The SVM that was trained for Pf protein antigen candidate prediction had three hyperparameters that needed to be optimized: the $k$-mer length, the positional uncertainty parameter $\sigma$, and the regularization parameter $C_{\text{SVM}}$. We performed a grid search utilizing repeated nested cross-validation to optimize all three hyperparameters. The resulting choices were $k = 1$, $\sigma = 18$, and $C_{\text{SVM}} = 0.001$.

Additionally, we used two more complex language embedding models that are commonly used for biological sequences: ESM-1b and ProtT5. The first, ESM-1b, is a pre-trained transformer model {\citep{rives2021biological}}, which is offered as a feature generator for downstream prediction models. It was developed to be used on biological sequences. ESM-1b follows the self-supervised bidirectional encoder representation from transformation (BERT) pre-training procedure. This language model is a transformer architecture with 33 layers and utilizes self-attention with 20 attention heads. The resulting features have a dimensionality of 1280 with a token context size of 1024. ESM-1b was trained on sequence clusters derived from the UniProt database {\citep{bairoch2005universal}}. We refer the interested reader to the original publication for all technical details about ESM-1b. The token context size together with a positional encoding of fixed length limits input sequences to a maximum of 1024 characters. Since there is a significant number of sequences in \textit{PlasmoFAB} that exceed this character limit, we followed published recommendations to cut the middle part of sequences that exceed the 1024 character limitation {\citep{thumuluri2022deeploc}} to be able to use ESM-1b on our benchmark. In total, 261 sequences were affected by this cutting procedure. The computed feature embeddings were used as inputs for the two tested downstream prediction models, LR and SVM. Again, we exclusively optimized the regularization parameters $C_{\text{LR}}$ and $C_{\text{SVM}}$, respectively. After performing the grid search, the optimal parameter choices were $C_{\text{LR}} = 0.15$ and $C_{\text{SVM}} = 20$.

The second language embedding that we used was ProtT5-XL-UniRef50 (ProtT5, \cite{elnaggar2020prottrans}). This transformer model, based on the language model T5 {\citep{raffel2020exploring}}, is specifically developed for biological data and prediction tasks. Similar to ESM-1b, ProtT5 acts as a feature generator for downstream prediction models. In contrast to other language models, ProtT5 follows an encoder-decoder approach and uses a simplified BERT training objective. The architecture employs 24 layers and also utilizes self-attention with 32 attention heads. ProtT5 has an embedding dimensionality of 1024. Since ProtT5 does not use a positional encoding of fixed length but learns a positional encoding for each attention head, the length of input sequences is not limited in theory. ProtT5 was first pre-trained on the BFD database {\citep{steinegger2019protein}} and fine-tuned on UniRef50 {\citep{suzek2015uniref}}. We refer the interested reader to the original publication for all technical details about ProtT5. Although sequence length is not limited when using ProtT5, finite computation power limits the usable sequence length in practice. With the computing resources available to us, an Nvidia Tesla V100 with 32GB RAM, the maximal usable sequence length was 6000 residues. Longer sequences were shortened in the same way we shortened sequences for ESM-1b. Five sequences in \textit{PlasmoFAB} were affected by this reduction of sequence length. Again, we used the feature embedding as inputs for the two downstream prediction models, LR and SVM, and optimized the regularization parameter via a grid search. The optimal parameters were $C_{\text{LR}} = 0.2$ and $C_{\text{SVM}} = 2.0$.

\subsection{Evaluating prediction models on \textit{PlasmoFAB}'s test sequences}

\begin{table*}[]
	\centering
	\caption[Test results on pfal dataset]{Performance of trained prediction models and prediction services on \textit{PlasmoFAB}'s test set. We trained different models on \textit{PlasmoFAB}'s training set including a support vector machine utilizing the oligo kernel ($\text{SVM}_{\text{oligo}}$), a combination of the a linear regression with either ESM1b or ProtT5 language model embedding ($\text{LR}_{\text{ESM1b}}$ and $\text{LR}_{\text{ProtT5}}$), and a support vector machine combined with either ESM1b or ProtT5 language model embedding ($\text{SVM}_{\text{ESM1b}}$ and $\text{SVM}_{\text{ProtT5}}$). Furthermore, we used publicly available, pre-trained prediction services on \textit{PlasmoFAB}'s test set. These services include Phobius, TMHMM, DeepTMHMM, Deeploc 1.0, and Deeploc 2.0.}
	\label{tab:pfal_test_res}
	\begin{tabular}{lllllll}
	    \toprule
		\textbf{Model} &  \textbf{MCC} & \textbf{F1} & \textbf{Bal. Acc.} & \textbf{Precision} & \textbf{Recall} & \textbf{Specificity}\\ 
		\midrule
		$\text{SVM}_{\text{oligo}}$ & 0.3145 & 0.5882 & 0.6500 & 0.7143 & 0.5000 & 0.8000 \\
		$\text{LR}_{\text{ESM1b}}$ & 0.7071 & 0.8000 & 0.8333 & \textbf{1.0000} & 0.6667 & \textbf{1.0000} \\
		$\text{SVM}_{\text{ESM1b}}$ & 0.7071 & 0.8000 & 0.8333 & \textbf{1.0000} & 0.6667 & \textbf{1.0000}\\
		$\text{LR}_{\text{ProtT5}}$ & \bf{0.7338} & \bf{0.8235} & \bf{0.8500} & \textbf{1.0000} & 0.7000 & \textbf{1.0000}\\
		$\text{SVM}_{\text{ProtT5}}$ & 0.6917 & 0.8077 & 0.8333 & 0.9545 & 0.7000 & 0.9666\\
		\\
		DeepTMHMM & 0.4395 & 0.6909 & 0.7167 & 0.7600 & 0.6333 & 0.8001\\
		DeepLoc 2.0 & 0.4009 & 0.7079 & 0.7009 & 0.6000 & 0.6923 & 0.7095 \\
		DeepLoc 1.0 & 0.2691 & 0.6071 & 0.6357 & 0.5667 & 0.6538 & 0.6176 \\
		TMHMM & 0.3015 & 0.6316 & 0.6500 & 0.6667 & 0.6000 & 0.7000 \\
		Phobius & 0.2722 & 0.6667 & 0.6333 & 0.6111 & \textbf{0.7333} & 0.5333 \\
		\bottomrule
	\end{tabular} 
\end{table*}

Table \ref{tab:pfal_test_res} shows the performance of all models on \textit{PlasmoFAB}'s test set. The models trained by ourselves can be directly applied to the test set. Since the publicly available prediction services do not always provide a binary output, we converted the prediction output for each service into a binary label. TMHMM and Phobius provide topology predictions for input sequences and we assigned a positive label to all samples with at least one predicted trans-membrane helix or at least one predicted extracellular region. Otherwise the sample was assigned a negative label. DeepTMHMM refines the prediction of TMHMM by providing a label for each residue in an input sample. For the DeepTMHMM output, we assigned a positive label to all samples with residues that had the membrane domain label ('\textit{M}') assigned. Furthermore, a positive label was assigned to samples where DeepTMHMM predicted the outside cell label ('\textit{O}') for all residues. If none of these conditions was fulfilled, the sample was assigned a negative label. DeepLoc 1.0 and 2.0 are tools for subcellular localization prediction and, hence, offer a multi-label output. Each label corresponds to a different subcellular localization. We used the top predicted label for each input sample. If this label was '\textit{cell membrane}' or '\textit{extracellular}', the sample was assigned a positive label, otherwise a negative label was assigned.

Our results show that models directly trained on \textit{PlasmoFAB} training set clearly outperform the available prediction services. The best performance was achieved by combining ProtT5 feature embedding with logistic regression. None of the tested prediction services was able to achieve a comparable performance to the specialized models.

\section{Discussion}
Computational antigen pre-screening with machine learning methods can drastically reduce time- and resource-consuming experimental exploration procedures and, thereby, accelerate development of drugs and vaccines. However, these computational pre-screening methods heavily depend on high-quality data to produce reliable results. In this work, we take important steps towards utilizing computational pre-screening for Malaria drug and vaccine development by providing \textit{PlasmoFAB}, a benchmark that consists of \textit{Plasmodium falciparum}-specific protein sequences with curated labels that distinguish between protein antigen candidates and intracellular proteins. 

Experimental validation is the gold standard to determine subcellular localization labels for proteins. We ensured that each label in PlasmoFAB achieves this gold standard or, if experimental validation is not feasible, comes as close to the gold standard as possible. As detailed in section \ref{sec:plasmofab}, the biggest subgroup of proteins that were assigned as antigen candidates was the group of VSA family members and known membrane proteins. We performed an exhaustive literature search and only included proteins into this subgroup for which published experimental evidence exists. Other subsets with experimentally validated labels are sporozoite proteins and proteins that contain the PEXEL/HT motif. Sporozoite proteins were validated by mass-spectrometry {\citep{swearingen2016interrogating}}. PEXEL/HT motif occurrence is a property of the protein sequence. This property is experimentally validated since \textit{PlasmoFAB} only includes experimentally validated protein sequences. Furthermore, there is experimental evidence that Pf parasites use the PEXEL/HT motif to export proteins {\citep{jonsdottir2021defining,osborne2010host}}. This supports our decision to include PEXEL/HT motif occurrence as an indication of protein antigen candidates. The last remaining subgroup in \textit{PlasmoFAB}'s positive set are proteins with known epitopes. IEDB only includes epitopes that are experimentally validated and we used BLAST to perform similarity matching between IEDB entries and Pf protein sequences. Although BLAST does not fulfill the gold standard of experimental validation, it is widely considered as the gold standard for sequence similarity matching. By restricting ourselves to BLAST matches with high or medium confidence, we ensured that the reduction in label quality of proteins in this subgroup is minimized. \textit{PlasmoFAB}'s negative set contains two groups of proteins: enzymes and intracellular proteins. We performed an exhaustive literature search to ensure that all included enzymes have experimental evidence of being intracellular. We excluded enzymes, if there is at least one publication with experimental evidence that suggests that the enzyme is being exported outside the cell. The other subgroup, intracellular proteins, were classified by a domain expert. While this does not fulfill the gold standard of experimental validation, we ensured to minimize the reduction in label quality by using domain expertise.

\textit{PlasmoFAB} uses data that belongs to the \textit{Plasmodium falciparum} strain 3D7. The genome of this specific strain of the Pf parasite was the first to be published by Gardner and colleagues in 2002 {\citep{gardner2002genome}}. It is still today one of the most important information sources for malaria research {\citep{olotu2013four,rts2015efficacy,mordmuller2017sterile,jagannathan2022malaria}}. Therefore, we made the decision to concentrate on Pf strain 3D7 for the first version of \textit{PlasmoFAB}. For future work, we want to further refine \textit{PlasmoFAB} by deriving high-quality labels for protein sequences of other Pf strains in order to incorporate as much information about Pf protein antigen candidates as possible into our benchmark.

One potentially surprising result is the sub-optimal performance of publicly available prediction services, like DeepTMHMM or DeepLoc 2.0, even though these services are relatively new and show impressive performance capabilities in their respective manuscripts. Our results do not provide evidence that the published performance capabilities of these models are overly optimistic or that they should not be used in general. On the contrary, we would like to emphasize that prediction services provide a fast and easy-to-use way for researchers without a strong background in machine learning to utilize prediction models in their research or the possibility to use prediction models even if not enough data for model training is available. However, our results highlight one common problem of general purpose models: their lack of out-of-distribution generalization {\citep{ye2021towards}}. Models learn certain aspects of the training data's distribution and allow trained models to achieve high prediction performance of unseen data as long as these data points came from the same distribution. However, if those unseen data points came from a different distribution, there is no guarantee that the model will be able to reliably make predictions on the new data. We see this out-of-distribution generalization issue in the relatively poor performance of the used prediction services. Since the Pf proteins are likely to be differently distributed than the proteins used to train the prediction services, these services perform poorly when applied to our test set. This result supports our claim that providing curated datasets with high-quality labels for model training is essential for maximising the potential of computational prediction methods on biological prediction tasks like the pre-screening of Pf protein antigen candidates. Therefore, our proposed \textit{PlasmoFAB} benchmark offers a solution to one fundamental obstacle in utilizing computational prediction methods in the development process of drugs and vaccines against malaria.

One goal of developing \textit{PlasmoFAB} was to provide the malaria research community with a tool to utilize machine learning in protein antigen exploration processes. However, the potential target user group of \textit{PlasmoFAB} can only benefit from the data if it fulfils two basic requirements. First, potential users have to be enabled to reliably find, access, and reuse data. And second, potential users have to be able to make an informed decision whether the data is applicable for their specific problem. We tackle the first problem by making \textit{PlasmoFAB} publicly available via Zenodo, which is a platform by researchers for researchers that aims to support open science. By uploading our dataset to Zenodo we ensure that the FAIR principles {\citep{wilkinson2016fair}} are taken into account. Additionally, we release \textit{PlasmoFAB} in form of comma-separated values (CSV) files. This file format is universally used in different research communities and should maximize the number of researchers that can use our dataset. Furthermore, we created a datasheet for \textit{PlasmoFAB} as described in {\citep{gebru2021datasheets}}. With this datasheet, we provide information about the motivation behind creating \textit{PlasmoFAB}, the creation process, the assumptions made, and applicable use cases. Users who are interested in using \textit{PlasmoFAB} can use the datasheet to make an informed decision about the applicability. 

\section{Conclusion}
With this work, we introduce \textit{PlasmoFAB}, a new and carefully curated benchmark for the training of models for \textit{Plasmodium falciparum} protein antigen candidate prediction. The benchmark was created by manually validating extracellular, surface-exposed, and intracellular Pf proteins to ensure high-quality labels for every sample in the dataset. Such a curated benchmark is an important prerequisite to incorporate learning models into pre-screening protocols for protein antigen candidates.

We furthermore compared commonly used prediction models with publicly available prediction services on the Pf protein antigen candidate prediction task. Our results show the limitations of existing prediction services, which are vastly outperformed by simpler prediction models that are specifically trained for Pf protein antigen candidate prediction. 

We are confident that our contribution provides a tool that can be used to help the research community to explore the vast number of \textit{Plasmodium falciparum} proteins with unknown functionality and identify new targets for drugs and vaccines against malaria.

\begin{ack}
Funded by the Deutsche Forschungsgemeinschaft (DFG, German Research Foundation) under Germany’s Excellence Strategy – EXC number 2064/1 – Project number 390727645. This research was supported by the German Federal Ministry of Education and Research (BMBF) project ’Training Center Machine Learning, Tübingen’ with grant number 01|S17054. This work was supported by the German Federal Ministry of Education and Research (BMBF): Tübingen AI Center, FKZ: 01IS18039A.
\end{ack}

\bibliographystyle{apalike}
\bibliography{references}

\begin{thebibliography}{}

\bibitem[AK et~al., 2021]{ak2021plasmodium}
AK, D., Shrivastava, D., Sahasrabuddhe, A.~A., Habib, S., and Trivedi, V.
  (2021).
\newblock Plasmodium falciparum fikk9. 1 is a monomeric serine-threonine
  protein kinase with features to exploit as a drug target.
\newblock {\em Chemical Biology \& Drug Design}.

\bibitem[Almagro~Armenteros et~al., 2017]{almagro2017deeploc}
Almagro~Armenteros, J.~J., S{\o}nderby, C.~K., S{\o}nderby, S.~K., Nielsen, H.,
  and Winther, O. (2017).
\newblock Deeploc: prediction of protein subcellular localization using deep
  learning.
\newblock {\em Bioinformatics}, 33(21):3387--3395.

\bibitem[Amos et~al., 2022]{amos2022veupathdb}
Amos, B., Aurrecoechea, C., Barba, M., Barreto, A., Basenko, E.~Y., Belnap, R.,
  Blevins, A.~S., B{\"o}hme, U., Brestelli, J., Brunk, B.~P., et~al. (2022).
\newblock Veupathdb: the eukaryotic pathogen, vector and host bioinformatics
  resource center.
\newblock {\em Nucleic Acids Research}, 50(D1):D898--D911.

\bibitem[Bairoch et~al., 2005]{bairoch2005universal}
Bairoch, A., Apweiler, R., Wu, C.~H., Barker, W.~C., Boeckmann, B., Ferro, S.,
  Gasteiger, E., Huang, H., Lopez, R., Magrane, M., et~al. (2005).
\newblock The universal protein resource (uniprot).
\newblock {\em Nucleic acids research}, 33(suppl\_1):D154--D159.

\bibitem[Baker, 2010]{baker2010making}
Baker, M. (2010).
\newblock Making membrane proteins for structures: a trillion tiny tweaks.
\newblock {\em Nature methods}, 7(6):429--434.

\bibitem[Chicco and Jurman, 2020]{chicco2020advantages}
Chicco, D. and Jurman, G. (2020).
\newblock The advantages of the matthews correlation coefficient (mcc) over f1
  score and accuracy in binary classification evaluation.
\newblock {\em BMC genomics}, 21(1):1--13.

\bibitem[Elnaggar et~al., 2020]{elnaggar2020prottrans}
Elnaggar, A., Heinzinger, M., Dallago, C., Rihawi, G., Wang, Y., Jones, L.,
  Gibbs, T., Feher, T., Angerer, C., Steinegger, M., et~al. (2020).
\newblock Prottrans: towards cracking the language of life's code through
  self-supervised deep learning and high performance computing.
\newblock {\em arXiv preprint arXiv:2007.06225}.

\bibitem[Gabler et~al., 2020]{gabler2020protein}
Gabler, F., Nam, S.-Z., Till, S., Mirdita, M., Steinegger, M., S{\"o}ding, J.,
  Lupas, A.~N., and Alva, V. (2020).
\newblock Protein sequence analysis using the mpi bioinformatics toolkit.
\newblock {\em Current Protocols in Bioinformatics}, 72(1):e108.

\bibitem[Gardner et~al., 2002]{gardner2002genome}
Gardner, M.~J., Hall, N., Fung, E., White, O., Berriman, M., Hyman, R.~W.,
  Carlton, J.~M., Pain, A., Nelson, K.~E., Bowman, S., et~al. (2002).
\newblock Genome sequence of the human malaria parasite plasmodium falciparum.
\newblock {\em Nature}, 419(6906):498--511.

\bibitem[Gebru et~al., 2021]{gebru2021datasheets}
Gebru, T., Morgenstern, J., Vecchione, B., Vaughan, J.~W., Wallach, H., Iii,
  H.~D., and Crawford, K. (2021).
\newblock Datasheets for datasets.
\newblock {\em Communications of the ACM}, 64(12):86--92.

\bibitem[Gupta et~al., 2015]{gupta2015conserved}
Gupta, A., Thiruvengadam, G., and Desai, S.~A. (2015).
\newblock The conserved clag multigene family of malaria parasites: essential
  roles in host--pathogen interaction.
\newblock {\em Drug Resistance Updates}, 18:47--54.

\bibitem[Hallgren et~al., 2022]{hallgren2022deeptmhmm}
Hallgren, J., Tsirigos, K.~D., Pedersen, M.~D., Armenteros, J. J.~A.,
  Marcatili, P., Nielsen, H., Krogh, A., and Winther, O. (2022).
\newblock Deeptmhmm predicts alpha and beta transmembrane proteins using deep
  neural networks.
\newblock {\em bioRxiv}.

\bibitem[Jagannathan and Kakuru, 2022]{jagannathan2022malaria}
Jagannathan, P. and Kakuru, A. (2022).
\newblock Malaria in 2022: Increasing challenges, cautious optimism.
\newblock {\em Nature communications}, 13(1):1--3.

\bibitem[Jonsdottir et~al., 2021]{jonsdottir2021defining}
Jonsdottir, T.~K., Gabriela, M., Crabb, B.~S., de~Koning-Ward, T.~F., and
  Gilson, P.~R. (2021).
\newblock Defining the essential exportome of the malaria parasite.
\newblock {\em Trends in Parasitology}, 37(7):664--675.

\bibitem[K{\"a}ll et~al., 2007]{kall2007advantages}
K{\"a}ll, L., Krogh, A., and Sonnhammer, E.~L. (2007).
\newblock Advantages of combined transmembrane topology and signal peptide
  prediction—the phobius web server.
\newblock {\em Nucleic acids research}, 35(suppl\_2):W429--W432.

\bibitem[Krogh et~al., 2001a]{krogh2001tmhmm}
Krogh, A., Larsson, B., Von~Heijne, G., and Sonnhammer, E.~L. (2001a).
\newblock Predicting transmembrane protein topology with a hidden markov model:
  application to complete genomes.
\newblock {\em Journal of molecular biology}, 305(3):567--580.

\bibitem[Krogh et~al., 2001b]{krogh2001predicting}
Krogh, A., Larsson, B., Von~Heijne, G., and Sonnhammer, E.~L. (2001b).
\newblock Predicting transmembrane protein topology with a hidden markov model:
  application to complete genomes.
\newblock {\em Journal of molecular biology}, 305(3):567--580.

\bibitem[Meinicke et~al., 2004]{meinicke2004oligo}
Meinicke, P., Tech, M., Morgenstern, B., and Merkl, R. (2004).
\newblock Oligo kernels for datamining on biological sequences: a case study on
  prokaryotic translation initiation sites.
\newblock {\em BMC bioinformatics}, 5(1):1--14.

\bibitem[Mordm{\"u}ller et~al., 2017]{mordmuller2017sterile}
Mordm{\"u}ller, B., Surat, G., Lagler, H., Chakravarty, S., Ishizuka, A.~S.,
  Lalremruata, A., Gmeiner, M., Campo, J.~J., Esen, M., Ruben, A.~J., et~al.
  (2017).
\newblock Sterile protection against human malaria by chemoattenuated pfspz
  vaccine.
\newblock {\em Nature}, 542(7642):445--449.

\bibitem[Obiero et~al., 2019]{obiero2019antibody}
Obiero, J.~M., Campo, J.~J., Scholzen, A., Randall, A., Bijker, E.~M.,
  Roestenberg, M., Hermsen, C.~C., Teng, A., Jain, A., Davies, D.~H., et~al.
  (2019).
\newblock Antibody biomarkers associated with sterile protection induced by
  controlled human malaria infection under chloroquine prophylaxis.
\newblock {\em Msphere}, 4(1):e00027--19.

\bibitem[Olotu et~al., 2013]{olotu2013four}
Olotu, A., Fegan, G., Wambua, J., Nyangweso, G., Awuondo, K.~O., Leach, A.,
  Lievens, M., Leboulleux, D., Njuguna, P., Peshu, N., et~al. (2013).
\newblock Four-year efficacy of rts, s/as01e and its interaction with malaria
  exposure.
\newblock {\em New England Journal of Medicine}, 368(12):1111--1120.

\bibitem[Osborne et~al., 2010]{osborne2010host}
Osborne, A.~R., Speicher, K.~D., Tamez, P.~A., Bhattacharjee, S., Speicher,
  D.~W., and Haldar, K. (2010).
\newblock The host targeting motif in exported plasmodium proteins is cleaved
  in the parasite endoplasmic reticulum.
\newblock {\em Molecular and biochemical parasitology}, 171(1):25--31.

\bibitem[Pedregosa et~al., 2011]{pedregosa2011scikit}
Pedregosa, F., Varoquaux, G., Gramfort, A., Michel, V., Thirion, B., Grisel,
  O., Blondel, M., Prettenhofer, P., Weiss, R., Dubourg, V., et~al. (2011).
\newblock Scikit-learn: Machine learning in python.
\newblock {\em the Journal of machine Learning research}, 12:2825--2830.

\bibitem[Raffel et~al., 2020]{raffel2020exploring}
Raffel, C., Shazeer, N., Roberts, A., Lee, K., Narang, S., Matena, M., Zhou,
  Y., Li, W., Liu, P.~J., et~al. (2020).
\newblock Exploring the limits of transfer learning with a unified text-to-text
  transformer.
\newblock {\em J. Mach. Learn. Res.}, 21(140):1--67.

\bibitem[Riley and Stewart, 2013]{riley2013immune}
Riley, E.~M. and Stewart, V.~A. (2013).
\newblock Immune mechanisms in malaria: new insights in vaccine development.
\newblock {\em Nature medicine}, 19(2):168.

\bibitem[Rives et~al., 2021]{rives2021biological}
Rives, A., Meier, J., Sercu, T., Goyal, S., Lin, Z., Liu, J., Guo, D., Ott, M.,
  Zitnick, C.~L., Ma, J., et~al. (2021).
\newblock Biological structure and function emerge from scaling unsupervised
  learning to 250 million protein sequences.
\newblock {\em Proceedings of the National Academy of Sciences},
  118(15):e2016239118.

\bibitem[Rts, 2015]{rts2015efficacy}
Rts, S. (2015).
\newblock Efficacy and safety of rts, s/as01 malaria vaccine with or without a
  booster dose in infants and children in africa: final results of a phase 3,
  individually randomised, controlled trial.
\newblock {\em The Lancet}, 386(9988):31--45.

\bibitem[Steinegger et~al., 2019]{steinegger2019protein}
Steinegger, M., Mirdita, M., and S{\"o}ding, J. (2019).
\newblock Protein-level assembly increases protein sequence recovery from
  metagenomic samples manyfold.
\newblock {\em Nature methods}, 16(7):603--606.

\bibitem[Suzek et~al., 2015]{suzek2015uniref}
Suzek, B.~E., Wang, Y., Huang, H., McGarvey, P.~B., Wu, C.~H., and Consortium,
  U. (2015).
\newblock Uniref clusters: a comprehensive and scalable alternative for
  improving sequence similarity searches.
\newblock {\em Bioinformatics}, 31(6):926--932.

\bibitem[Swearingen et~al., 2016]{swearingen2016interrogating}
Swearingen, K.~E., Lindner, S.~E., Shi, L., Shears, M.~J., Harupa, A., Hopp,
  C.~S., Vaughan, A.~M., Springer, T.~A., Moritz, R.~L., Kappe, S.~H., et~al.
  (2016).
\newblock Interrogating the plasmodium sporozoite surface: identification of
  surface-exposed proteins and demonstration of glycosylation on csp and trap
  by mass spectrometry-based proteomics.
\newblock {\em PLoS pathogens}, 12(4):e1005606.

\bibitem[Taha and Hanbury, 2015]{taha2015metrics}
Taha, A.~A. and Hanbury, A. (2015).
\newblock Metrics for evaluating 3d medical image segmentation: analysis,
  selection, and tool.
\newblock {\em BMC medical imaging}, 15(1):1--28.

\bibitem[Tarr et~al., 2014]{tarr2014conserved}
Tarr, S.~J., Moon, R.~W., Hardege, I., and Osborne, A.~R. (2014).
\newblock A conserved domain targets exported phistb family proteins to the
  periphery of plasmodium infected erythrocytes.
\newblock {\em Molecular and biochemical parasitology}, 196(1):29--40.

\bibitem[Thumuluri et~al., 2022]{thumuluri2022deeploc}
Thumuluri, V., Almagro~Armenteros, J.~J., Johansen, A.~R., Nielsen, H., and
  Winther, O. (2022).
\newblock Deeploc 2.0: multi-label subcellular localization prediction using
  protein language models.
\newblock {\em Nucleic Acids Research}.

\bibitem[Tuteja, 2007]{tuteja2007malaria}
Tuteja, R. (2007).
\newblock Malaria- an overview.
\newblock {\em The FEBS journal}, 274(18):4670--4679.

\bibitem[Vita et~al., 2019]{vita2019immune}
Vita, R., Mahajan, S., Overton, J.~A., Dhanda, S.~K., Martini, S., Cantrell,
  J.~R., Wheeler, D.~K., Sette, A., and Peters, B. (2019).
\newblock The immune epitope database (iedb): 2018 update.
\newblock {\em Nucleic acids research}, 47(D1):D339--D343.

\bibitem[Wahlgren et~al., 2017]{wahlgren2017variant}
Wahlgren, M., Goel, S., and Akhouri, R.~R. (2017).
\newblock Variant surface antigens of plasmodium falciparum and their roles in
  severe malaria.
\newblock {\em Nature Reviews Microbiology}, 15(8):479--491.

\bibitem[WHO et~al., 2021]{WHO2021}
WHO et~al. (2021).
\newblock {\em World malaria report 2021}.
\newblock World Health Organization.

\bibitem[WHO et~al., 2022]{world2022world}
WHO et~al. (2022).
\newblock {\em World malaria report 2022}.
\newblock World Health Organization.

\bibitem[Wilkinson et~al., 2016]{wilkinson2016fair}
Wilkinson, M.~D., Dumontier, M., Aalbersberg, I.~J., Appleton, G., Axton, M.,
  Baak, A., Blomberg, N., Boiten, J.-W., da~Silva~Santos, L.~B., Bourne, P.~E.,
  et~al. (2016).
\newblock The fair guiding principles for scientific data management and
  stewardship.
\newblock {\em Scientific data}, 3(1):1--9.

\bibitem[Wu, 2019]{wu2019evaluation}
Wu, H.~M. (2019).
\newblock Evaluation of the sick returned traveler.
\newblock In {\em Seminars in diagnostic pathology}, volume~36, pages 197--202.
  Elsevier.

\bibitem[Ye et~al., 2021]{ye2021towards}
Ye, H., Xie, C., Cai, T., Li, R., Li, Z., and Wang, L. (2021).
\newblock Towards a theoretical framework of out-of-distribution
  generalization.
\newblock {\em Advances in Neural Information Processing Systems},
  34:23519--23531.

\bibitem[Zimmermann et~al., 2018]{zimmermann2018completely}
Zimmermann, L., Stephens, A., Nam, S.-Z., Rau, D., K{\"u}bler, J., Lozajic, M.,
  Gabler, F., S{\"o}ding, J., Lupas, A.~N., and Alva, V. (2018).
\newblock A completely reimplemented mpi bioinformatics toolkit with a new
  hhpred server at its core.
\newblock {\em Journal of molecular biology}, 430(15):2237--2243.

\end{thebibliography}

\end{document}